# Nonradiative energy transfer between thickness-controlled halide perovskite nanoplatelets

*Andreas Singldinger\*, Moritz Gramlich, Christoph Gruber, Carola Lampe, and Alexander S. Urban\**

Nanospectroscopy Group and Center for Nanoscience (CeNS), Nano-Institute Munich, Department of Physics, Ludwig-Maximilians-Universität München, Königinstr. 10, 80539 Munich, Germany

**Corresponding Authors**

A.S.: andreas.singldinger@physik.uni-muenchen.de

A.S.U.: urban@lmu.de

ABSTRACT

Despite showing great promise for optoelectronics, the commercialization of halide perovskite nanostructure-based devices is hampered by inefficient electrical excitation and strong exciton binding energies. While transport of excitons in an energy-tailored system via Förster resonance energy transfer (FRET) could be an efficient alternative, halide ion migration makes the realization of cascaded structures difficult. Here, we show how these could be obtained by exploiting the pronounced quantum confinement effect in two-dimensional $CsPbBr_3$-based nanoplatelets (NPls). In thin films of NPls of two predetermined thicknesses, we observe an enhanced acceptor photoluminescence (PL) emission and a decreased donor PL lifetime. This indicates a FRET-mediated process, benefitted by the structural parameters of the NPls. We determine corresponding transfer rates up to $k_{FRET} = 0.99 \, ns^{-1}$ and efficiencies of nearly $\eta_{FRET} = 70 \, \%$. We also show FRET to occur between perovskite NPls of other thicknesses. Consequently, this strategy could lead to tailored, enery cascade nanostructures for improved optoelectronic devices.

**TOC GRAPHICS**

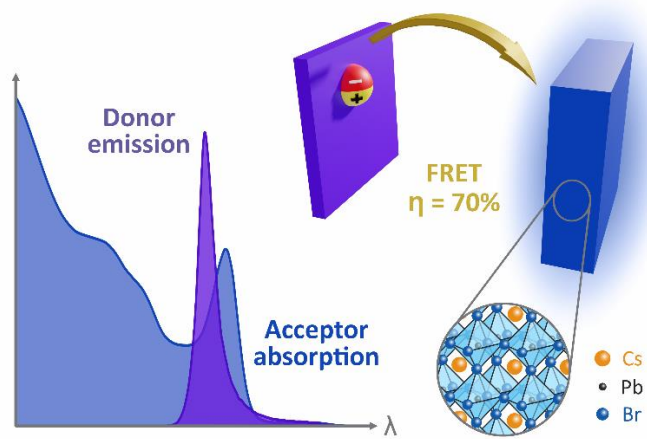

MAIN TEXT

Halide perovskites are one of the hottest semiconducting materials for optoelectronic applications due to a range of fascinating properties.[1, 2] With bandgaps tunable throughout the visible range,[3] high absorption cross-sections,[4] and photoluminescence (PL) quantum yields approaching 100%,[5] potential uses range from solar cells[6, 7] and photodetectors[8] to light-emitting diodes (LEDs)[9, 10] and lasers[11] and even more exotic applications such as gamma-ray detectors[12] and remote thermometers.[13] With the initial focus on fabricating large-grain thin films for photovoltaics, fabrication has since spread to two-dimensional (2D) perovskite phases and nanocrystals (NCs).[14-17] These possess an additional tuning mechanism, as shrinking any dimension below the excitonic Bohr radius induces quantum confinement, strongly affecting the optoelectronic properties.[18] Examples of this have been demonstrated for 2D nanoplatelets (NPls) and nanosheets, 1D nanowires and nanorods, and for 0D quantum dots (QDs).[2, 19] Halide perovskites represent an interesting conundrum, as these marvelous properties are counterbalanced by a number of partially significant limitations. Instability to moisture, heat and strong light exposure,[20, 21] migration of ions during operation of devices[22] and the (current) necessity of containing lead[23] are some of the issues currently impeding commercialization. Ion migration, especially, has presented a problem, as it prevents the formation of heterostructures and thereupon reliant strategies such as energy funnels.[24, 25] A possible workaround could be the exploitation of quantum confinement in perovskite nanostructures, such as NPls, whose absorption and emission properties can be tuned over wide ranges by controlling their thickness.[26, 27] Previous studies on NPls of different materials have already shown efficient energy transfer mechanisms due to large spectral overlap and efficient lateral ordering in thin films.[28, 29] This energy transfer was shown to occur highly efficiently by the Förster resonance energy transfer (FRET) mechanism even outpacing detrimental Auger recombination. The FRET process relies on a spectral overlap of the emission of a donor material and absorption of an acceptor material, the orientation of their dipole moments and a small separation between them. Such an energy transfer was shown in perovskite thin films with varying grain thicknesses and in inhomogeneously broadened NC aggregates, but to date, not on separate monodisperse NCs.[30-32] Herein, we demonstrate the first such occurrence, using previously gained expertise on precisely controlling the thickness of $CsPbBr_3$ NPls.[33] Mixing colloidal dispersions of 2 and 3 monolayer (ML) thick NPls, we first show an enhanced PL emission of the thicker NPls, which act as acceptors. Proof of a nonradiative energy transfer mechanism comes via time-resolved PL measurements in thin films comprising mixtures of 2 and 3 ML NPls. We show an increasing PL acceptor lifetime and a simultaneous decrease of the donor lifetime as the molar acceptor:donor ratio ($A{:}D$) of the NPls is varied. With only a narrow ligand spacer between adjacent NPls, the process is highly efficient ($\eta_{FRET} \approx 70\,\%$) and extremely fast ($k_{FRET} \approx 1\,ns^{-1}$). We also demonstrate that energy transfer is prevalent in NPl mixtures of other thicknesses. These results open the way for highly defined cascaded energy transfer structures for realizing efficient energy harvesting, high optical gain and even electrically-pumped lasing.

Halide perovskite NPls were synthesized according to a modified version of our previously reported method (see Experimental Methods for details).[33] In this past work, the main goal was to boost the radiative efficiency of the NPls, which we achieved by applying a passivating step after the synthesis. A drawback of this process is that the NPls tend to be unstable and especially in films, often rapidly grow thicker, as evidenced by a progressive redshift of the PL maxima. For the energy transfer measurements, however, it was critical that we produce thin films comprising exactly two distinct NPl thicknesses, otherwise, the lifetimes would be skewed and produce unreliable values.

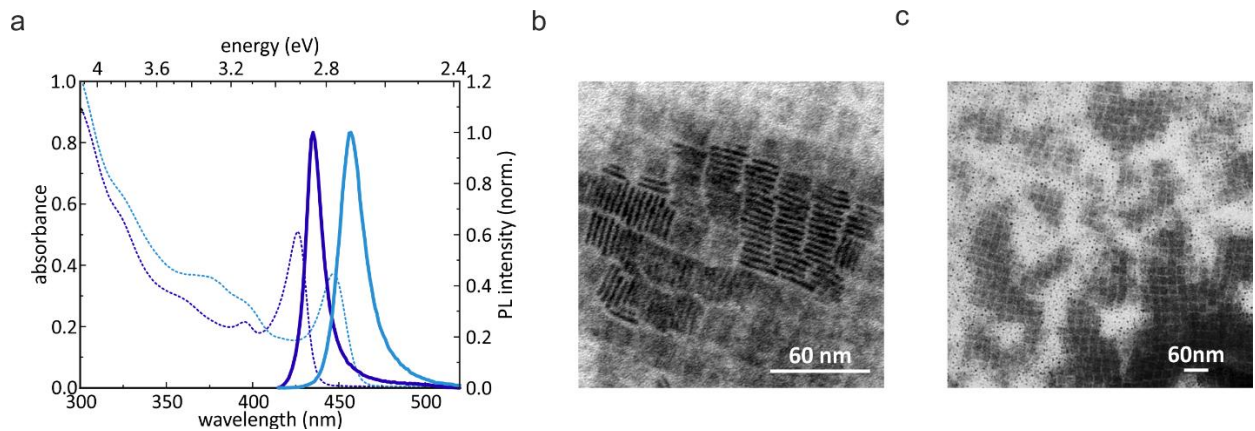

**Figure 1.** Thickness-controlled halide perovskite nanoplatelets (NPls). a) Absorption (dashed lines) and PL (solid lines) spectra of 2 ML (purple) and 3 ML (blue) NPls. b,c) TEM images of 2 ML NPls from which a thickness of $1.3 \pm 0.2 \ nm$ and side lengths of $19 \pm 1 \ nm$ are determined.

Accordingly, we modified the synthesis to obtain less enhanced, but very stable NPl dispersions. The most essential parameter for this was the applied volume of enhancement solution, as it turns out that the balance between $PbBr_2$ and ligands in the dispersions is crucial for the long-term stability of the NPls. For this study, we focused on NPls of two thicknesses, 2 MLs and 3 MLs. To assess the quality of the two individual dispersions we acquired UV-VIS and PL spectra (Figure 1a), which reveal sharp excitonic absorption peaks and single narrow PL peaks at 435 nm and 457 nm, for the 2 and 3 ML NPls respectively, matching our previous results.[33] Importantly, there is a strong spectral overlap between the emission spectrum of the 2 ML and the absorption spectrum of the 3 ML sample. This is an essential prerequisite for realizing a donor/acceptor energy transfer system. According to this spectral alignment, the 2 ML NPls appear suitable as donors, while the 3 ML NPls can accept the energy from the donor system. Any subsequent reference to donor/acceptor in this manuscript refers to 2 ML/3 ML NPls. The narrow PL peaks with full widths at half maximum (fwhm) of 76 meV and 104 meV and the small Stokes shifts (60 meV and 61 meV) are indicative of the monodispersity of the dispersions. Further evidence for this is provided through transmission electron microscopy (TEM) images, as shown for the 2 ML NPls (Figure 1b,c). The 2 ML NPls possess a quadratic shape with a side length of 19 ± 1 nm and a thickness of 1.3 ± 0.2 nm, matching the reported lattice constant of 0.59 nm for $CsPbBr_3$.[34] The 3 ML NPls are also quadratic, but slightly larger with a side length of 32 ± 1 nm and a thickness of 1.9 ± 0.2 nm (see Supporting Information, Figure S1). Often self-assembling into long stacks during the drying process, the NPls passivated with long alkyl chain ligands are separated by 2.3 nm and 2.6 nm for the 2 ML and 3 ML samples, respectively. This arrangement and the small distance are important for enabling energy transfer processes based on dipole-dipole interaction.[29]

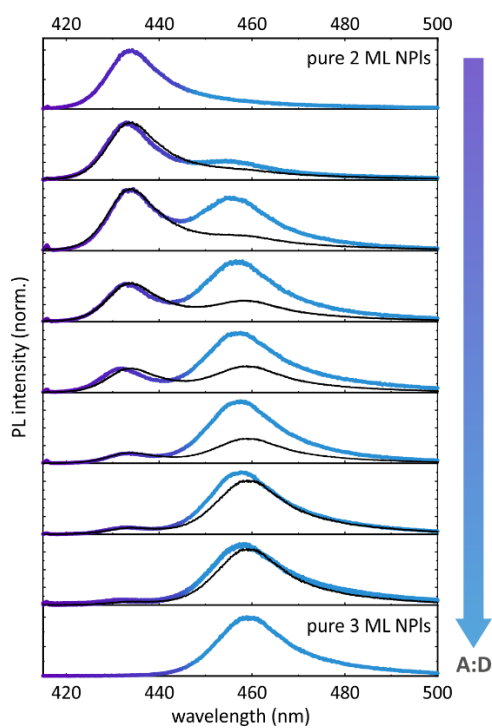

**Figure 2.** Energy transfer between NPls in thin films. PL spectra (colored lines) of mixed 2 ML and 3 ML NPl thin films obtained via drop-casting and ranging from pure 2 ML (top) to pure 3 ML NPls (bottom). Shown in black are the calculated PL spectra obtained by adding the pure 2 ML and

3 ML spectra in the mixing ratio of the respective species. The increase in intensity of the 3 ML emission peak represents a clear sign of energy transfer between the two NPl types.

To investigate possible energy transfer processes, we mixed the two NPl dispersions in fixed molar ratios ($A{:}D$). To this end, we initially determined the absorption cross-sections at 400 nm for the individual NPls and used this in combination with their optical density to calculate the concentration of the dispersions, which we found to be $c_{2\,ML} = 1.20 \cdot 10^{14}\ cm^{-3}$ and $c_{3\,ML} = 2.47 \cdot 10^{14}\ cm^{-3}$, for the 2 and 3 ML NPls respectively (see Supporting Information for details). UV-VIS spectra of the mixtures were then acquired and compared to calculated spectra obtained from a weighted addition of the individual NPl spectra (see Figure S2). Clearly, these two match very closely demonstrating that the absorption of the mixture resembles a simple superposition of the two individual absorption spectra. More importantly, the two NPl species are not affected by the mixing process. PL spectra of the mixtures, however, exhibit a different situation, as for all mixtures containing a significant amount of both NPl species, the PL emission of the donor is decreased while that of the acceptor is increased (see Figure S3). This suggests that energy transfer already occurs in the mixtures, however, at these NPl concentrations, this is likely due to photons being emitted by the donor and reabsorbed by the acceptor NPls.[35] Thus, solid-state films were prepared by drop-casting these mixtures onto $SiO_2$-coated silicon substrates and PL emission spectra of these films were acquired (Figure 2, colored curves). To investigate whether energy transfer occurs, we again calculated the expected PL spectra for the case that no energy transfer was present according to the molar ratios of the donor/acceptor NPls (black curves). Here, the curves are normalized to the donor emission peak for clarity. Clearly, in all cases the PL emission of the acceptor is enhanced, again pointing to energy transfer between the NPls. While the drop-casted films are considerably thinner than the cuvettes used for the dispersion, making radiative energy transfer less likely, these results do not exclude reabsorption as the transfer mechanism. For this, time-resolved PL measurements are necessary.

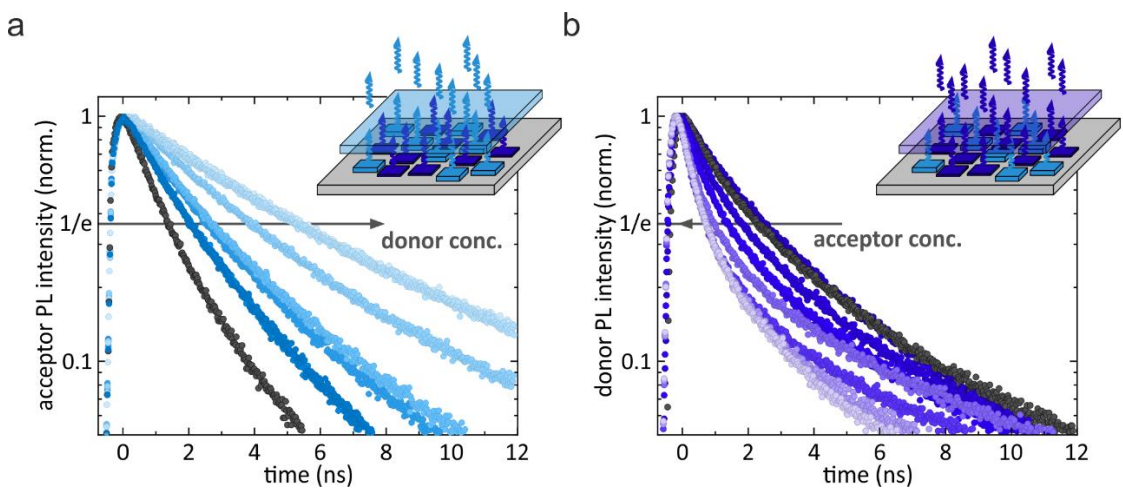

**Figure 3.** FRET-modified PL lifetimes in mixed-NPl thin films. The PL decays of the pure 2 and 3 ML NPls are represented by the black curves. a) PL decay curves of 3 ML NPls show increasing lifetimes as the molar donor concentration is increased (dark to light coloring). b) The corresponding PL decays of 2 ML NPls show diminishing lifetimes for increasing acceptor molar ratios. These two trends show the nonradiative nature of the energy transfer process in the films. The inset shows a schematic of the experimental setup with the filtered emission from the mixed NPl thin films.

Accordingly, we employed a time-correlated single-photon counting (TCSPC) setup to investigate the PL decay and corresponding lifetimes of the donor and acceptor NPls in the solid-state heterostructures. While the donor and acceptor PL emission maxima are easily discernible from one another, as can be seen in Figure 1a, there is a slight overlap of the two spectra. However, for the lifetime measurements, it is crucial to ensure that only the PL emission from either the donor or the acceptor reaches the detector, which is difficult for the more unbalanced mixtures, due to this overlap. To achieve this, we installed two separate edge filters for the 2 ML and 3 ML samples (see Figure S4a). The PL of the 2 ML and 3 ML samples decays within several nanoseconds, exhibiting a multiexponential nature, possibly due to different dielectric surroundings or due to bright and dark subpopulations (see black curves in Figure 3a,b).[33] We thus extract the times at which the PL intensity has dropped to 1/e of its original value as a measure for the PL lifetimes, obtaining donor and acceptor lifetimes $\tau_D = 2.26\ ns$ and $\tau_A = 1.45\ ns$, respectively. In the mixed films, this acceptor lifetime increases up to a maximum value of $\tau_{AD} = 5.36\ ns$ as progressively more donor is added (Figure 3a). This is a clear indicator of an energy transfer process, with a feeding of the acceptors leading to a slower PL decay. We were not able to observe a delayed onset of the acceptor emission, likely due to the fact we were exciting both samples simultaneously, with the excitation wavelength located above the continuum onset of the two NPl systems. As before, the change in the acceptor lifetime does not define the actual transfer mechanism. Therefore, we investigate the PL lifetime of the donor in the mixed samples (Figure 3b). Here, the opposite trend to the acceptor is visible, with the decay accelerating up to $\tau_{DA} = 0.7\ ns$ as the molar acceptor concentration increases. The fact that the donor emission becomes faster means that an additional decay pathway must be present, as simple reabsorption of photons by the acceptor would not change the donor lifetime. Hence, we postulate that FRET occurs from the donor to acceptor NPls as previously observed for CdSe NPl films.[28, 29]

To quantify the efficiency of the energy transfer process, we calculate the FRET rates $k_{FRET}$ and FRET efficiencies $\eta_{FRET}$ from the donor PL lifetimes in the pure and mixed samples according to [36]

$$k_{FRET} = \frac{1}{\tau_{DA}} - \frac{1}{\tau_D} \quad and \quad \eta_{FRET} = 1 - \frac{\tau_{DA}}{\tau_D}.$$

These two physical quantities are plotted against the $A{:}D$ ratio (Figure 4). For the smallest amount of acceptor present ($A{:}D = 0.11$) both the FRET rate and FRET efficiency assume low values of $k_{FRET} = 0.12\ ns^{-1}$ and $\eta_{FRET} = 22\ \%$. As the amount of acceptor is increased, both values rise rapidly. The FRET efficiency saturates at a ratio $A{:}D = 20$, reaching a maximum of 69 %. The FRET rate increase slows down for the highest $A{:}D$ ratios, but still does not seem completely saturated with a maximally determined value of $k_{FRET} = 0.99\ ns^{-1}$. Both behaviors make sense, as a higher amount of acceptor will increase the likelihood of a donor NPl having acceptor NPls in its proximity to which it can transfer its energy. However, at one point all donor NPls will obtain their maximal transfer probabilities. These are limited by the NPl spacing given by the organic ligand layers and by the overlap of emission and absorption spectra of the donor and acceptor, respectively. A further increase of both values might be possible by using shorter passivating ligands or by enhancing the order in the films, as stacks of NPls should be favorable for a more efficient dipole-dipole interaction. Nevertheless, the values are comparable to those reported recently for a similar CdSe-NPl system.[37]

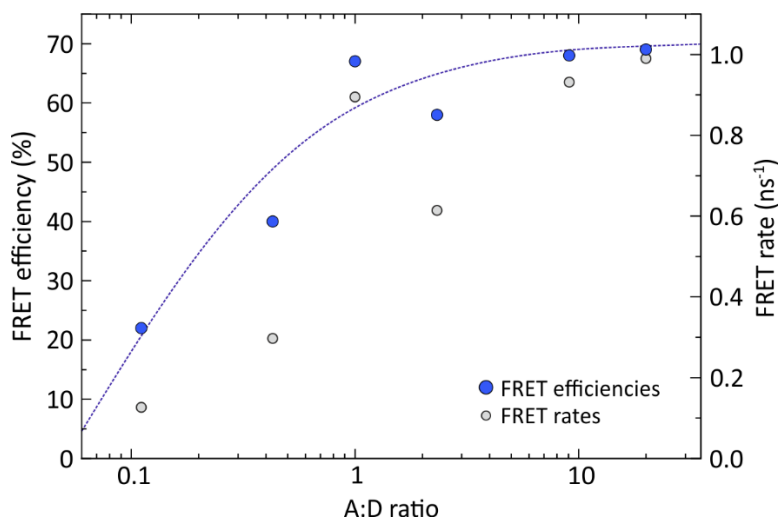

**Figure 4.** The extracted donor PL lifetimes obtained from the mixed NPl thin films were used to calculate both the FRET efficiency (purple circles) and the FRET rates (gray circles). Both values grow with increasing $A:D$ ratio, saturating for large acceptor concentrations.

To confirm the energy transfer process, we also synthesized NPls of increased thickness, with 4 and 6 ML (see Figure S4b). The PL emission from these two samples peaks at 477 nm and 503 nm, respectively. The latter, as evidenced by a broader PL spectrum (124 meV fwhm) is clearly not monodisperse, and both samples suffered from stability issues upon mixing with other NPl dispersions. Consequently, we cannot use these to determine the FRET rates and efficiencies between NPls of specific thicknesses. However, we can use them to confirm that nonradiative energy transfer also occurs in these systems. Accordingly, we performed a similar set of measurements for 2 ML and 4 ML mixtures and for 3 ML and 6 ML mixtures (see Figures S5, S6). In both sets, we were again able to observe a slowing of the acceptor decay accompanied by an accelerating donor decay, again indicative of a FRET-mediated process. The FRET rates in all cases tended to be similar while the efficiencies for these latter systems peaked at 49% and 52%, respectively (see Tables S1, S2). These lower values could be due to the aforementioned issues or due to less efficient processes. More work will need to be done to optimize the syntheses and stabilize the NPls to elucidate this question.

In summary, we have presented the first demonstration of FRET between two defined halide perovskite nanocrystal populations. For this we used $CsPbBr_3$-based NPls with a precisely adjusted thickness, as these have strongly thickness-dependent energy levels, enabling nonradiative energy transfer. By fabricating thin films comprising different $A:D$ ratios of NPls, we show a progressive lengthening of the acceptor PL lifetime and a shortening of the donor lifetime, indicative of the FRET mechanism. We use the reduced donor lifetimes in the mixtures to estimate the efficiencies and transfer rates of the process. Both increase strongly with increasing $A:D$ ratio, reaching maximum values of $\eta_{FRET} = 69\,\%$ and $k_{FRET} = 0.99\,ns^{-1}$. The FRET process is not limited to these thicknesses, as we show for two other thickness combinations. With energy transfer between perovskite NCs typically impossible due to halide ion exchange, size-controllable NCs present an interesting alternative for realizing cascaded energy transfer structures. However, the efficiency of the NCs and most importantly their stability must be improved for this and for achieving optoelectronic integration.

ASSOCIATED CONTENT

**Supporting Information Available:** Experimental section, materials and methods for the synthesis, concentration calculation of NPl dispersions, TEM characterization and optical characterization including UV-VIS, PL spectroscopy and TCSPC measurements.

AUTHOR INFORMATION

*Email: andreas.singldinger@physik.uni-muenchen.de (A.S.)

*Email: urban@lmu.de (A.S.U.)

Website: www.nanospec.de

Twitter: @NanospecGroup

Instagram: @nanospecgroup

**ORCID**

Andreas Singldinger: 0000-0002-3613-6570

Moritz Gramlich: 0000-0002-4733-4708

Carola Lampe: 0000-0001-7833-7306

Alexander S. Urban: 0000-0001-6168-2509

**Notes**

The authors declare no competing financial interest.

ACKNOWLEDGMENTS

We gratefully acknowledge support by the Bavarian State Ministry of Science, Research and Arts through the grant "Solar Technologies go Hybrid (SolTech)" and by the Deutsche Forschungsgemeinsschaft (DFG) under Germany's Excellence Strategy EXC 2089/1-390776260. This work was also supported by the European Research Council Horizon 2020 through the ERC Grant Agreement PINNACLE (759744). Furthermore, we thank Matías Herran and Stefan Maier aiding in the absorption measurements of our samples.

REFERENCES

(1)  Manser, J. S.; Christians, J. A.; Kamat, P. V., Intriguing Optoelectronic Properties of Metal Halide Perovskites. *Chem. Rev.* **2016,** *116*, 12956-13008.

(2)  Shamsi, J.; Urban, A. S.; Imran, M.; De Trizio, L.; Manna, L., Metal Halide Perovskite Nanocrystals: Synthesis, Post-Synthesis Modifications, and Their Optical Properties. *Chem. Rev.* **2019,** *119*, 3296-3348.

(3)  Protesescu, L.; Yakunin, S.; Bodnarchuk, M. I.; Krieg, F.; Caputo, R.; Hendon, C. H.; Yang, R. X.; Walsh, A.; Kovalenko, M. V., Nanocrystals of Cesium Lead Halide Perovskites ($CsPbX_3$, X = Cl, Br, and I): Novel Optoelectronic Materials Showing Bright Emission with Wide Color Gamut. *Nano Lett.* **2015,** *15*, 3692-3696.

(4)  Fakharuddin, A.; Schmidt-Mende, L. Hybrid Organic/Inorganic and Perovskite Solar Cells. In *Molecular Devices for Solar Energy Conversion and Storage*; **2018**; pp 187-227.

(5) Liu, F.; Zhang, Y.; Ding, C.; Kobayashi, S.; Izuishi, T.; Nakazawa, N.; Toyoda, T.; Ohta, T.; Hayase, S.; Minemoto, T.; Yoshino, K.; Dai, S.; Shen, Q., Highly Luminescent Phase-Stable CsPbI3 Perovskite Quantum Dots Achieving Near 100% Absolute Photoluminescence Quantum Yield. *ACS Nano* **2017,** *11*, 10373-10383.

(6) Park, N.-G.; Grätzel, M.; Miyasaka, T.; Zhu, K.; Emery, K., Towards stable and commercially available perovskite solar cells. *Nat. Energy* **2016,** *1,* 16152.

(7) Saliba, M.; Matsui, T.; Seo, J.-Y.; Domanski, K.; Correa-Baena, J.-P.; Nazeeruddin, M. K.; Zakeeruddin, S. M.; Tress, W.; Abate, A.; Hagfeldt, A.; Grätzel, M., Cesium-containing triple cation perovskite solar cells: improved stability, reproducibility and high efficiency. *Energy Environ. Sci.* **2016,** *9*, 1989-1997.

(8) Leung, S.-F.; Ho, K.-T.; Kung, P.-K.; Hsiao, V. K. S.; Alshareef, H. N.; Wang, Z. L.; He, J.-H., A Self-Powered and Flexible Organometallic Halide Perovskite Photodetector with Very High Detectivity. *Adv. Mater.* **2018,** *30*, 1704611.

(9) Lin, K.; Xing, J.; Quan, L. N.; de Arquer, F. P. G.; Gong, X.; Lu, J.; Xie, L.; Zhao, W.; Zhang, D.; Yan, C.; Li, W.; Liu, X.; Lu, Y.; Kirman, J.; Sargent, E. H.; Xiong, Q.; Wei, Z., Perovskite light-emitting diodes with external quantum efficiency exceeding 20 per cent. *Nature* **2018,** *562*, 245-248.

(10) Choi, Y. J.; Debbichi, L.; Lee, D.-K.; Park, N.-G.; Kim, H.; Kim, D., Light Emission Enhancement by Tuning the Structural Phase of APbBr3 (A = CH3NH3, Cs) Perovskites. *The Journal of Physical Chemistry Letters* **2019,** *10*, 2135-2142.

(11) Zhu, H.; Fu, Y.; Meng, F.; Wu, X.; Gong, Z.; Ding, Q.; Gustafsson, M. V.; Trinh, M. T.; Jin, S.; Zhu, X., Lead halide perovskite nanowire lasers with low lasing thresholds and high quality factors. *Nat. Mater.* **2015,** *14*, 636-642.

(12) He, Y.; Matei, L.; Jung, H. J.; McCall, K. M.; Chen, M.; Stoumpos, C. C.; Liu, Z.; Peters, J. A.; Chung, D. Y.; Wessels, B. W.; Wasielewski, M. R.; Dravid, V. P.; Burger, A.; Kanatzidis, M. G., High spectral resolution of gamma-rays at room temperature by perovskite CsPbBr3 single crystals. *Nat. Commun.* **2018,** *9*, 1609.

(13) Yakunin, S.; Benin, B. M.; Shynkarenko, Y.; Nazarenko, O.; Bodnarchuk, M. I.; Dirin, D. N.; Hofer, C.; Cattaneo, S.; Kovalenko, M. V., High-resolution remote thermometry and thermography using luminescent low-dimensional tin-halide perovskites. *Nat. Mater.* **2019,** *18*, 846-852.

(14) Tsai, H.; Nie, W.; Blancon, J.-C.; Stoumpos, C. C.; Soe, C. M. M.; Yoo, J.; Crochet, J.; Tretiak, S.; Even, J.; Sadhanala, A.; Azzellino, G.; Brenes, R.; Ajayan, P. M.; Bulović, V.; Stranks, S. D.; Friend, R. H.; Kanatzidis, M. G.; Mohite, A. D., Stable Light-Emitting Diodes Using Phase-Pure Ruddlesden-Popper Layered Perovskites. *Adv. Mater.* **2018,** *30*, 1704217.

(15) Cao, D. H.; Stoumpos, C. C.; Farha, O. K.; Hupp, J. T.; Kanatzidis, M. G., 2D Homologous Perovskites as Light-Absorbing Materials for Solar Cell Applications. *J. Am. Chem. Soc.* **2015,** *137*, 7843-7850.

(16) Swarnkar, A.; Chulliyil, R.; Ravi, V. K.; Irfanullah, M.; Chowdhury, A.; Nag, A., Colloidal CsPbBr3Perovskite Nanocrystals: Luminescence beyond Traditional Quantum Dots. *Angew. Chem. Int. Ed.* **2015,** *54*, 15424-15428.

(17) Congreve, D. N.; Weidman, M. C.; Seitz, M.; Paritmongkol, W.; Dahod, N. S.; Tisdale, W. A., Tunable Light-Emitting Diodes Utilizing Quantum-Confined Layered Perovskite Emitters. *ACS Photonics* **2017,** *4*, 476-481.

(18) Fu, Y.; Zheng, W.; Wang, X.; Hautzinger, M. P.; Pan, D.; Dang, L.; Wright, J. C.; Pan, A.; Jin, S., Multicolor Heterostructures of Two-Dimensional Layered Halide Perovskites that Show Interlayer Energy Transfer. *Journal of the American Chemical Society* **2018,** *140*, 15675-15683.

(19) Akkerman, Q. A.; Raino, G.; Kovalenko, M. V.; Manna, L., Genesis, challenges and opportunities for colloidal lead halide perovskite nanocrystals. *Nat. Mater.* **2018,** *17*, 394-405.

(20) Berhe, T. A.; Su, W.-N.; Chen, C.-H.; Pan, C.-J.; Cheng, J.-H.; Chen, H.-M.; Tsai, M.-C.; Chen, L.-Y.; Dubale, A. A.; Hwang, B.-J., Organometal halide perovskite solar cells: degradation and stability. *Energy Environ. Sci.* **2016,** *9*, 323-356.

(21) Conings, B.; Drijkoningen, J.; Gauquelin, N.; Babayigit, A.; D'Haen, J.; D'Olieslaeger, L.; Ethirajan, A.; Verbeeck, J.; Manca, J.; Mosconi, E.; Angelis, F. D.; Boyen, H.-G., Intrinsic Thermal Instability of Methylammonium Lead Trihalide Perovskite. *Adv. Energy Mater.* **2015,** *5*, 1500477.

(22) Azpiroz, J. M.; Mosconi, E.; Bisquert, J.; De Angelis, F., Defect migration in methylammonium lead iodide and its role in perovskite solar cell operation. *Energy Environ. Sci.* **2015,** *8*, 2118-2127.

(23) Giustino, F.; Snaith, H. J., Toward Lead-Free Perovskite Solar Cells. *ACS Energy Lett.* **2016,** *1*, 1233-1240.

(24) Palazon, F.; Akkerman, Q. A.; Prato, M.; Manna, L., X-ray Lithography on Perovskite Nanocrystals Films: From Patterning with Anion-Exchange Reactions to Enhanced Stability in Air and Water. *ACS Nano* **2016,** *10*, 1224-1230.

(25) Hintermayr, V. A.; Lampe, C.; Low, M.; Roemer, J.; Vanderlinden, W.; Gramlich, M.; Bohm, A. X.; Sattler, C.; Nickel, B.; Lohmuller, T.; Urban, A. S., Polymer Nanoreactors Shield Perovskite Nanocrystals from Degradation. *Nano Lett.* **2019,** *19*, 4928-4933.

(26) Hintermayr, V. A.; Richter, A. F.; Ehrat, F.; Döblinger, M.; Vanderlinden, W.; Sichert, J. A.; Tong, Y.; Polavarapu, L.; Feldmann, J.; Urban, A. S., Tuning the Optical Properties of Perovskite Nanoplatelets through Composition and Thickness by Ligand-Assisted Exfoliation. *Adv. Mater.* **2016,** *28*, 9478-9485.

(27) Sichert, J. A.; Tong, Y.; Mutz, N.; Vollmer, M.; Fischer, S.; Milowska, K. Z.; García Cortadella, R.; Nickel, B.; Cardenas-Daw, C.; Stolarczyk, J. K.; Urban, A. S.; Feldmann, J., Quantum Size Effect in Organometal Halide Perovskite Nanoplatelets. *Nano Lett.* **2015,** *15*, 6521-6527.

(28) Guzelturk, B.; Olutas, M.; Delikanli, S.; Kelestemur, Y.; Erdem, O.; Demir, H. V., Nonradiative energy transfer in colloidal CdSe nanoplatelet films. *Nanoscale* **2015,** *7*, 2545-2551.

(29) Rowland, C. E.; Fedin, I.; Zhang, H.; Gray, S. K.; Govorov, A. O.; Talapin, D. V.; Schaller, R. D., Picosecond energy transfer and multiexciton transfer outpaces Auger recombination in binary CdSe nanoplatelet solids. *Nat. Mater.* **2015,** *14*, 484-489.

(30) Yuan, M.; Quan, L. N.; Comin, R.; Walters, G.; Sabatini, R.; Voznyy, O.; Hoogland, S.; Zhao, Y.; Beauregard, E. M.; Kanjanaboos, P.; Lu, Z.; Kim, D. H.; Sargent, E. H., Perovskite energy funnels for efficient light-emitting diodes. *Nat. Nanotech.* **2016**, *11*, 872-877.

(31) de Weerd, C.; Gomez, L.; Zhang, H.; Buma, W. J.; Nedelcu, G.; Kovalenko, M. V.; Gregorkiewicz, T., Energy Transfer between Inorganic Perovskite Nanocrystals. *J. Phys. Chem. C* **2016**, *120*, 13310-13315.

(32) Yantara, N.; Bruno, A.; Iqbal, A.; Jamaludin, N. F.; Soci, C.; Mhaisalkar, S.; Mathews, N., Designing Efficient Energy Funneling Kinetics in Ruddlesden-Popper Perovskites for High-Performance Light-Emitting Diodes. *Adv. Mater.* **2018**, *30*, 1800818.

(33) Bohn, B. J.; Tong, Y.; Gramlich, M.; Lai, M. L.; Döblinger, M.; Wang, K.; Hoye, R. L. Z.; Müller-Buschbaum, P.; Stranks, S. D.; Urban, A. S.; Polavarapu, L.; Feldmann, J., Boosting Tunable Blue Luminescence of Halide Perovskite Nanoplatelets through Postsynthetic Surface Trap Repair. *Nano Lett.* **2018**, *18*, 5231-5238.

(34) Tong, Y.; Bladt, E.; Aygüler, M. F.; Manzi, A.; Milowska, K. Z.; Hintermayr, V. A.; Docampo, P.; Bals, S.; Urban, A. S.; Polavarapu, L.; Feldmann, J., Highly Luminescent Cesium Lead Halide Perovskite Nanocrystals with Tunable Composition and Thickness by Ultrasonication. *Angew. Chem. Int. Ed.* **2016**, *55*, 13887-13892.

(35) Davis, N. J. L. K.; de la Peña, F. J.; Tabachnyk, M.; Richter, J. M.; Lamboll, R. D.; Booker, E. P.; Wisnivesky Rocca Rivarola, F.; Griffiths, J. T.; Ducati, C.; Menke, S. M.; Deschler, F.; Greenham, N. C., Photon Reabsorption in Mixed $CsPbCl_3:CsPbI_3$ Perovskite Nanocrystal Films for Light-Emitting Diodes. *J. Phys. Chem. C* **2017**, *121*, 3790-3796.

(36) Majoul, I.; Jia, Y.; Duden, R. Practical fluorescence resonance energy transfer or molecular nanobioscopy of living cells. In *Handbook Of Biological Confocal Microscopy*; Pawley, J. B., Ed.; Springer: Boston, MA, USA, **2006**; pp 788-808.

(37) Yu, J.; Sharma, M.; Delikanli, S.; Birowosuto, M. D.; Demir, H. V.; Dang, C., Mutual Energy Transfer in a Binary Colloidal Quantum Well Complex. *J. Phys. Chem. Lett.* **2019,** *10*, 5193-5199.